\documentclass[12pt,english,a4paper]{article}
\usepackage[T1]{fontenc}
\usepackage[latin9]{inputenc}
\usepackage{amsmath}
\usepackage{graphicx}

\makeatletter
\usepackage{float}

\usepackage{epsfig}

\makeatother

\usepackage{babel}

\begin{document}

\makeatother

\title{\textbf{\large Quantum gravity corrections to gauge theories with
a cutoff regularization}}

\author{G. Cynolter and E. Lendvai}

\date{MTA-ELTE Research Group in Theoretical Physics, E\"otv\"os University,
Budapest, Hungary }
\maketitle
\begin{abstract}
The gravitational waves recently observed by the LIGO collaboration
is an experimental evidence that the weak field approximation of general
relativity is a viable, calculable scenario. As a non-renormalizable
theory, gravity can be successfully considered as an effective quantum
field theory with reliable, but limited predictions. Though the influence
of gravity on gauge and other interactions of elementary particles
is still an open question. In this chapter we calculate the lowest
order quantum gravity contributions to the QED beta function in an
effective field theory picture with a momentum cutoff. We use a recently
proposed 4 dimensional improved momentum cutoff that preserves gauge
and Lorentz symmetries. We find that there is a non-vanishing quadratic
contribution to the photon 2-point function but after renormalization
that does not lead to the running of the original coupling. We comment
on corrections to the other gauge interactions and Yukawa couplings
of heavy fermions. We argue that gravity cannot turn gauge interactions
asymptotically free.
\end{abstract}
%\preprint{ITP-Budapest 670?}

\section{Introduction}

Recently, in the latest four-five years there were two outstanding
discoveries in the area of physics of fundamental interactions. The
upgraded LIGO experiment observed \cite{ligo} gravitational waves
in 2015 and published in 2016 and the LHC has announced the discovery
of the Higgs boson in Run I in 2012. The observation of the gravitational
waves traveling with the speed of light is a direct evidence that
the weak field approximation of general relativity can be used reliably
in high precision calculation. Furthermore the source of the event
GW150914 is found to be consistent with merging of two black hole
with mass approximately 39 and 32 solar masses and the LIGO collaboration
found no evidence for violations of general relativity in this strong
field regime of gravity. Despite this success perturbatively quantized
general relativity is still considered to be a non-renormalizable
theory due to its dimensionful coupling constant $\kappa$ with negative
mass dimension ($\kappa^{2}=32\pi G_{\mathrm{N}}=1/M_{P}^{2}$). This
way the naively quantized Eintein theory cannot be considered as a
fundamental theory at the quantum level \cite{veltmanh} as newer
and newer counter terms have to be introduced at each order of the
perturbative calculation and the cutoff cannot be taken to infinity.
However Donoghue argued that assuming there is some yet unknown, well
defined theory of quantum gravity that yields the observed general
relativity as a low energy limit, then the Einstein-Hilbert action
can be used to calculate gravitational correction in the framework
of effective field theories (well) below the Planck mass $M_{P}\simeq1.2\times10^{18}$
GeV\cite{greff,greff2}. The subject was reviewed in details by Burgess
in \cite{burg}.

The other important recent achievement was the discovery of the SM
(Standard Model) Higgs boson with a mass approximately 125 GeV by
the ATLAS and CMS collaborations \cite{atlas,cms}. So far the properties
of the 125 GeV scalar are in complete agreement with the SM Higgs
predictions, few sigma anomalies in the photon-photon and the lepton
number violating mu-tau final states (at CMS) have disappeared. This
value of the Higgs mass falls in a special region where not only several
different decay channel are experimentally tested, but it implies
that the SM is perturbatively renormalizable up to $M_{P}$. The complete
Standard Model might be valid up to the Planck scale \cite{stabil,misa}.
In this case we live close to the stability region in the $\left(m_{top},\: M_{H}\right)$
plane in a metastable world \cite{stab3}, where the tunneling to
the lower, real minimum is longer than the lifetime of our Universe.
Considering the SM or its extensions valid up to the Planck scale
gravity can influence the SM observables and running parameters at
the loop-level. The gravitational corrections can be estimated in
an effective field theory framework and may be important as they may
modify the running of the various coupling, possibly alter the gauge
coupling unification and the conclusions concerning the stability
of the Standard Model. In the seventies the first attempts using dimensional
regularization showed that only higher order operators get renormalized
at one-loop order \cite{deser}.

The effective field theory treatment of gravity was recently used
to study quantum corrections to gauge and other theories. In the pioneering
work, starting the new era, Robinson and Wilczek argued that the gravity
contribution to the Yang-Mills beta function is quadratically divergent
and negative, further the corrections point toward asymptotic freedom
\cite{robinson}. There were several controversial results about this
claim in the literature. Pietrykowski showed in \cite{pietr} that
in the Maxwell-Einstein theory the result is gauge dependent and doubted
the validity of the Robinson Wilczek result. Toms repeated the calculation
in the gauge choice independent background field method using dimensional
regularization and has found no quantum gravity contribution to the
beta function \cite{toms1}. Diagrammatic calculation employing dimensional
regularization and naive momentum cutoff \cite{rodigast} found vanishing
quadratic contribution. The authors showed that the logarithmic divergences
renormalize the dimension-6 operators in agreement with the early
results of Deser et al. \cite{deser}. Toms later applied proper time
cutoff regularization and claimed that the quadratic dependence on
the energy remains in the QED one-loop effective action \cite{toms2}.
Analysis using the background field method employing the gauge invariant
Vilkovisky-DeWitt formalism \cite{wu,pietr2,he} and special loop
regularization that respects Ward identities both found non-vanishing
quadratic contributions to the beta function, but \cite{wu} with
sign opposite to \cite{robinson,toms2}. Nielsen showed that the quadratic
divergences are generally still gauge dependent in the Vilkovisky-DeWitt
formalism \cite{nielsen}. In the asymptotic safety scenario \cite{weinberg,reuter}
Reuter et al. has found going beyond naive perturbation theory that
gravity contribution points towards asymptotic freedom of the Yang-Mills
theory \cite{reuter2}, later Litim et al. showed that gravity does
not contribute to the running of the gauge coupling \cite{litim}.
In a higher derivative renormalizable theory of gravity the authors
\cite{narain} showed that the gravity correction vanishes in any
gauge theory. There are many various results (for more complete list
see the references in e.g. \cite{pietr2}), sometimes contradicting
to each other and the physical reality of quadratic corrections to
the gauge coupling was questioned \cite{ellis,don1,brazil,grcutoff}.
The situation could be clarified using a straightforward cutoff calculation
respecting the symmetries of the models and correctly interpreting
the divergences appearing in the calculations.

Earlier the present authors developed a new improved momentum cutoff
regularization which by construction respects the gauge and Lorentz
symmetries of gauge theories at one loop level \cite{uj}. In this
chapter we discuss the application to the effective Maxwell-Einstein
and Einstein-Yang-Mills systems to estimate the regularized gravitational
corrections to the photon/gluon two and three point functions in the
simplest possible model and later discuss more involved theories.

The paper is organized as follows. In section 2. the effective gravity
contribution to quantum electrodynamics is calculated, in section
3. the renormalization is discussed. In chapter 4 corrections to a
Yang-Mills theory is presented. The paper is closed with conclusions
and an appendix summarizing the improved momentum cutoff method.

\section{Effective Maxwell-Einstein theory}

In this section we present the calculation of the gravitational quantum
corrections to the photon self energy in the simple Einstein-Maxwell
theory, given by the Lagrangian \cite{grcutoff}

\begin{equation}
S=\int\mathrm{d}^{4}x\sqrt{-g}\left[\frac{2}{\kappa^{2}}R-\frac{1}{2}g^{\mu\nu}g^{\alpha\beta}F_{\mu\nu}F_{\alpha\beta}\right],\label{eq:EH}\end{equation}
where $R$ is the Ricci scalar, $\kappa^{2}=32\pi G_{\mathrm{N}}$
and $F_{\mu\nu}$ denotes the $U(1)$ field strength tensor. Quantum
effects are calculated in the weak field expansion around the flat
Minkowski metric ($\eta_{\mu\nu}=(1,-1,-1,-1)$)

\begin{equation}
g_{\mu\nu}=\eta_{\mu\nu}+\kappa h_{\mu\nu}(x).\label{eq:lin1}\end{equation}
This is considered an exact relation, but the inverse of the metric
contains higher order terms \begin{equation}
g^{\mu\nu}=\eta^{\mu\nu}-\kappa h^{\mu\nu}+\kappa^{2}h_{\alpha}^{\mu}h^{\nu\alpha}+\ldots\,,\label{eq:lin2}\end{equation}
in an effective treatment it can be truncated at the second order.
The photon propagator is defined in the Landau gauge\[
\frac{g_{\mu\nu}-\frac{k_{\mu}k_{\nu}}{k^{2}}}{k^{2}-i\epsilon},\]
 and the graviton propagator in de Donder, or harmonic gauge, where
the gauge condition is (with $h=h_{\alpha}^{\alpha}$)\begin{equation}
\partial^{\nu}h_{\mu\nu}-\frac{1}{2}\partial_{\mu}h=0.\label{eq:gauge}\end{equation}
Expanding the Lagrangian up to second order in the graviton field
we get the following graviton propagator in $d$ dimensions \begin{equation}
G_{\alpha\beta\gamma\delta}^{G}(k)=i\frac{\frac{1}{2}\eta_{\alpha\gamma}\eta_{\beta\delta}+\frac{1}{2}\eta_{\alpha\delta}\eta_{\beta\gamma}-\frac{1}{d-2}\eta_{\alpha\beta}\eta_{\gamma\delta}}{k^{2}-i\epsilon}.\label{eq:propg}\end{equation}
There are two relevant vertices with two photons. The two photon-graviton
vertex is\begin{eqnarray}
V_{\gamma\gamma G}(k_{1\mu},k_{2\nu},\alpha,\beta) & \!=\! & -i\frac{\kappa}{2}\left[\eta_{\alpha\beta}\left(k_{1\nu}k_{2\mu}-\eta_{\mu\nu}(k_{1}k_{2})\right)+\right.\nonumber \\
 &  & \!\left.+Q_{\mu\nu,\alpha\beta}(k_{1}k_{2})+Q_{k_{1}k_{2},\alpha\beta}\eta_{\mu\nu}-Q_{\mu k_{2},\alpha\beta}k_{1\nu}-Q_{k_{1}\nu,\alpha\beta}k_{2\mu}\right],\label{eq:V3}\end{eqnarray}
and the two photon-two graviton vertex is even more complicated

\begin{eqnarray}
V_{\gamma\gamma GG}(k_{1\mu},k_{2\nu},\alpha,\beta,\gamma,\delta) & \!=\! & -i\frac{\kappa^{2}}{4}\left[P_{\alpha\beta\gamma\delta}\left(k_{1\nu}k_{2\mu}-\eta_{\mu\nu}(k_{1}k_{2})\right)+U_{\mu\nu,\alpha\beta,\gamma\delta}(k_{1}k_{2})+\phantom{\kappa^{2}}\right.\nonumber \\
 &  & \!+U_{k_{1}k_{2},\alpha\beta,\gamma\delta}\eta_{\mu\nu}-U_{\mu k_{2},\alpha\beta,\gamma\delta}k_{1\nu}-U_{k_{1}\nu,\alpha\beta,\gamma\delta}k_{2\mu}+\nonumber \\
 &  & \!+Q_{\mu\nu,\alpha\beta}Q_{\gamma\delta,k_{1}k_{2}}+Q_{\mu\nu,\gamma\delta}Q_{\alpha\beta,k_{1}k_{2}}\nonumber \\
 &  & \!\left.-Q_{k_{1}\nu,\alpha\beta}Q_{\mu k_{2},\gamma\delta}-Q_{\mu k_{2},\alpha\beta}Q_{k_{1}\nu,\gamma\delta}\phantom{\kappa^{2}}\right].\label{eq:V4}\end{eqnarray}
For the sake of simplicity we have defined \begin{equation}
U_{\mu\nu,\alpha\beta,\gamma\delta}=\eta_{\mu\alpha}P_{\nu\beta,\gamma\delta}+\eta_{\mu\beta}P_{\alpha\nu,\gamma\delta}+\eta_{\mu\gamma}P_{\alpha\beta,\nu\delta}+\eta_{\mu\delta}P_{\alpha\beta,\gamma\nu},\label{eq:UU}\end{equation}
\begin{equation}
P_{\alpha\beta,\mu\nu}=\eta_{\mu\alpha}\eta_{\nu\beta}+\eta_{\mu\beta}\eta_{\nu\alpha}-\eta_{\mu\nu}\eta_{\alpha\beta},\label{eq:PP}\end{equation}
and finally\begin{equation}
Q_{\alpha\beta,\mu\nu}=\eta_{\mu\alpha}\eta_{\nu\beta}+\eta_{\mu\beta}\eta_{\nu\alpha}.\label{eq:QQ}\end{equation}

There are two graphs contributing to the photon self energy with two
vertices \eqref{eq:V3gg} giving $\Pi^{(a)}$ (Fig. 1. left) and one
4-leg vertex \eqref{eq:V4} providing $\Pi^{(b)}$ (Fig. 1. right).
We calculated the finite and divergent parts of the 2-point function
with improved cutoff, naive 4-dimensional momentum cutoff and dimensional
regularization. The improved momentum cutoff is defined to respect
gauge and Lorentz symmetries and allows for shifting the loop momentum
under divergent loop-integrals. Compared to naive cutoff it changes
the coefficient of the quadratic divergence and gives a finite shift
in the presence of a universal logarithmic divergence. The details
of the new regularization scheme with some example and outlook on
the broad literature can be found in the Appendix. For comparison,
using the technique of dimensional regularization with different assumptions
about treating the number of dimensions $d$ in the propagator and
vertices various quadratically divergent cutoff results can be identified
using the connection between cutoff and dimenisonal regularization
results, see \eqref{eq: quad} in the Appendix. Each of these calculation
defines a different regularization scheme.

\begin{figure}
\begin{centering}
(a)\includegraphics[height=3cm]{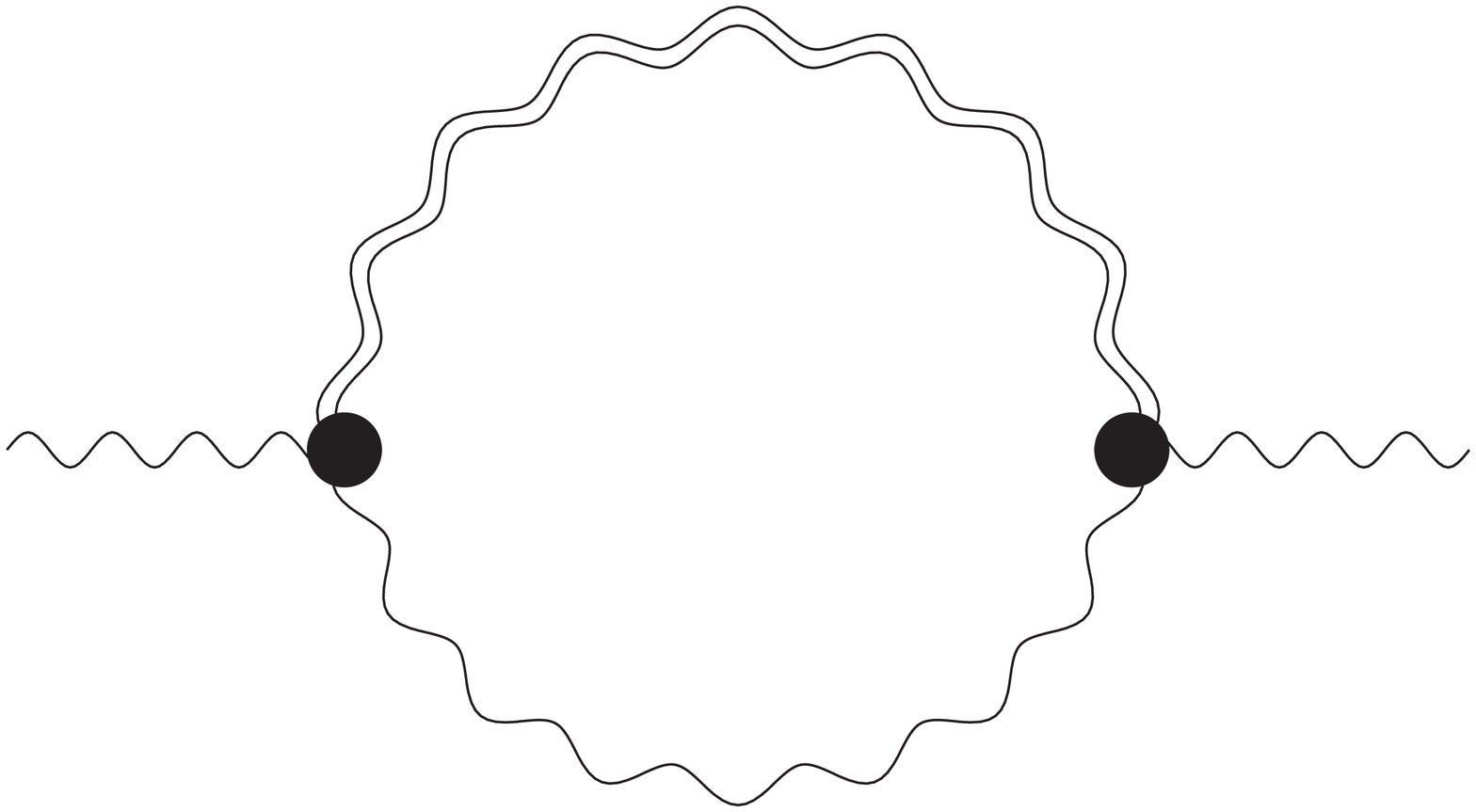}(b)\includegraphics[height=3cm]{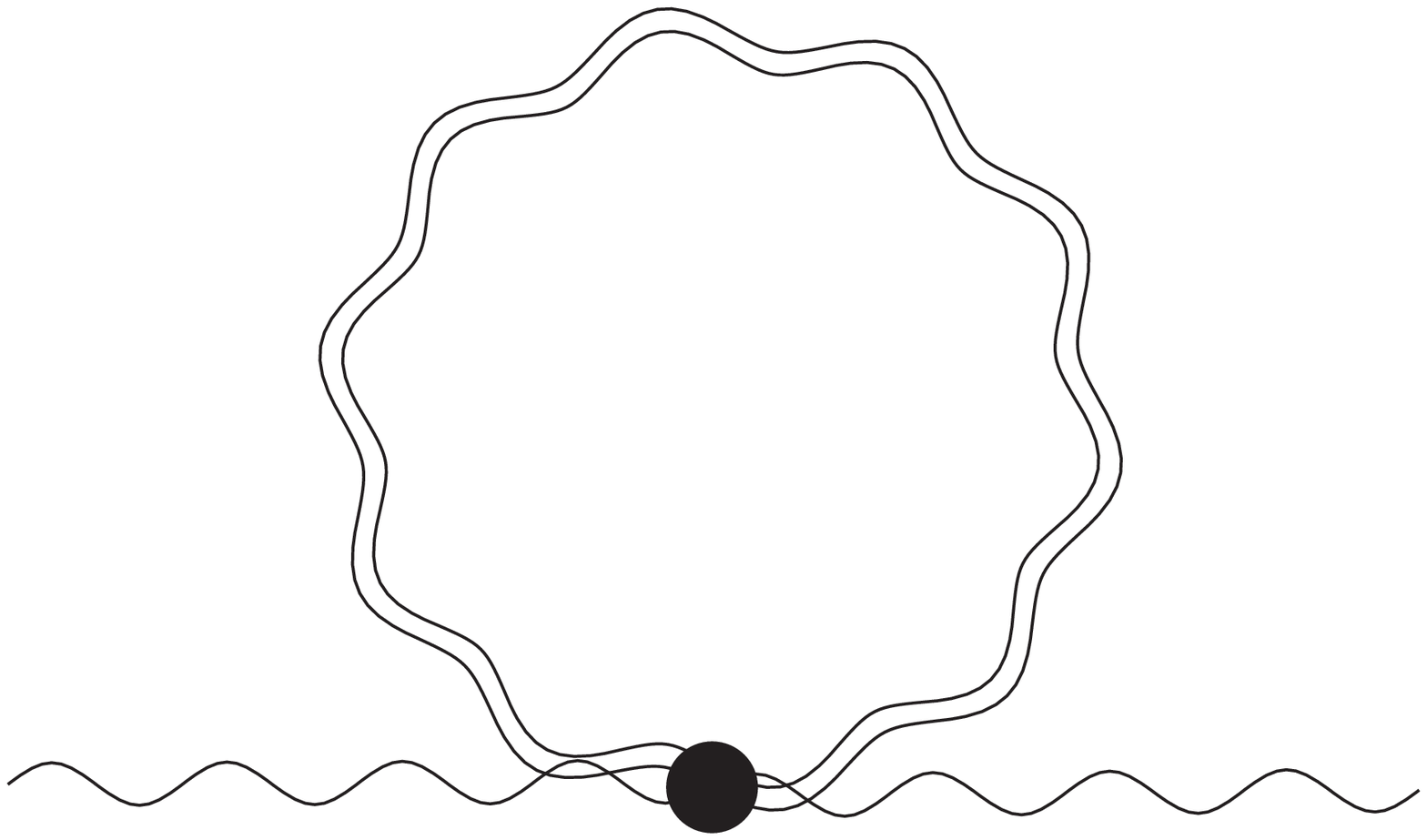}
\par\end{centering}

\centering{}\caption{Feynman graphs with graviton (double) lines contributing to the photon
two point function.}

\end{figure}

The calculation of the diagrams is straightforward, we used the symbolic
manipulation program \textit{FORM} \cite{form} to deal with the large
number of terms. The quadratically divergent contributions of the
two graphs with \textbf{improved cutoff }(I) do not cancel each other
\begin{eqnarray}
\Pi_{\mu\nu}^{\mathrm{I}(a)}(p) & = & \frac{i}{16\pi^{2}}\kappa^{2}\left(p^{2}\eta_{\mu\nu}-p_{\mu}p_{\nu}\right)\left(-2\Lambda^{2}-\frac{1}{6}p^{2}\left(\ln\left(\frac{\Lambda^{2}}{p^{2}}\right)+\frac{2}{3}\right)\right),\label{eq:impi1}\\
\Pi_{\mu\nu}^{\mathrm{I}(b)}(p) & = & \frac{i}{16\pi^{2}}\kappa^{2}\left(p^{2}\eta_{\mu\nu}-p_{\mu}p_{\nu}\right)\left(\phantom{2}\frac{3}{2}\Lambda^{2}\right).\label{eq:impi2}\end{eqnarray}

In the \textbf{naive cutoff} (N) calculation using \eqref{eq:negyed}
there is a cancellation of the $\Lambda^{2}$ terms, the finite term
do not match the previous one, and it is remarkable that the result
is transverse without any subtractions

\begin{eqnarray}
\Pi_{\mu\nu}^{\mathrm{N}(a)}(p) & = & \frac{i}{16\pi^{2}}\kappa^{2}\left(p^{2}\eta_{\mu\nu}-p_{\mu}p_{\nu}\right)\left(-\frac{3}{2}\Lambda^{2}-\frac{1}{6}p^{2}\ln\left(\frac{\Lambda^{2}}{p^{2}}\right)-\frac{7}{36}p^{2}\right),\label{eq:cupi1}\\
\Pi_{\mu\nu}^{\mathrm{N}(b)}(p) & = & \frac{i}{16\pi^{2}}\kappa^{2}\left(p^{2}\eta_{\mu\nu}-p_{\mu}p_{\nu}\right)\left(\phantom{-}\frac{3}{2}\Lambda^{2}\right).\label{eq:cupi2}\end{eqnarray}

In \textbf{dimensional regularization} (DR) the space-time dimension
is continued in all terms originating from the gauge and gravitational
part, too (e.g. $\eta_{\mu}^{\mu}=d=4-2\epsilon$). The result (just
as using the naive cutoff above) agrees with \cite{rodigast} (without
the finite terms which are first given here) 

\begin{eqnarray}
\Pi_{\mu\nu}^{DR1(a)}(p) & = & \frac{i}{16\pi^{2}}\kappa^{2}\left(p^{2}\eta_{\mu\nu}-p_{\mu}p_{\nu}\right)\left(-\frac{1}{6}p^{2}\left(\frac{2}{\epsilon}+\ln\left(\frac{\mu^{2}}{p^{2}}\right)+\frac{1}{6}\right)\right),\label{eq:DRpi1}\\
\Pi_{\mu\nu}^{DR1(b)}(p) & = & 0,\label{eq:DRpi2}\end{eqnarray}
where we have omitted the constants $-\gamma_{E}+\ln4\pi$ beside
$2/\epsilon$. 

In what follows we present various {}``cutoff'' results we arrived
at using the technique of dimensional regularization based on different
assumptions about the continuation of the dimension. Each result defines
a different regularization scheme, and they are denoted by the superscript
$DR1,\, DR2,\, DR3$ and the corresponding cutoff results by $\Lambda1,\,\Lambda2,\,\Lambda3$
based on the extension of dimensional regularization.

Now with the help of the equations in the appendix \eqref{eq: quad},
\eqref{eq:log} and \eqref{eq:finite} we can define three cutoff
results based on the dimensional regularization one. In the first
case the dimension is modified in each terms where $d$ appears, also
in the graviton propagator \eqref{eq:propg}, though gravity is not
a dynamical theory in $d=2$. Each graph is quadratically divergent,
even $1/(\epsilon-1)^{2}$ type of singularities appear in single
graphs, but they cancel in the sum of the graphs, like the $\frac{1}{\epsilon^{2}}$
terms in usual gauge theories (e.g. in QCD) at two loops.\begin{equation}
\Pi_{\mu\nu}^{\mathrm{\Lambda1}}(p)=\frac{i}{16\pi^{2}}\kappa^{2}\left(p^{2}\eta_{\mu\nu}-p_{\mu}p_{\nu}\right)\left(-\frac{1}{4}\Lambda^{2}-\frac{1}{6}p^{2}\left(\ln\left(\frac{\Lambda^{2}}{p^{2}}\right)-\frac{5}{6}\right)\right)\label{eq:L1}\end{equation}
also quadratically divergent, but only the coefficient of the logarithmic
term agrees with other results.

To find connection with existing, partially controversial literature,
we have performed the calculation with weaker assumptions. First the
term in the graviton propagator is set $\frac{1}{d-2}=\frac{1}{2}$
as is usually done in earlier results e.g. \cite{don1,brazil}. The
divergent part of the dimensional regularization result agrees with
\cite{rodigast}. The contribution of the tadpole in Fig. 1b $\Pi^{\mathrm{DR2(b)}}$
vanishes, the sum is\begin{equation}
\Pi_{\mu\nu}^{DR2}(p)=\frac{i}{16\pi^{2}}\kappa^{2}\left(p^{2}\eta_{\mu\nu}-p_{\mu}p_{\nu}\right)\left(-\frac{1}{6}p^{2}\left(\frac{2}{\epsilon}+\ln\left(\frac{\mu^{2}}{p^{2}}\right)+\frac{1}{6}\right)\right).\label{eq:DR2}\end{equation}
We can identify a cutoff result, Fig. 1b gives $\Pi_{\mu\nu}^{\mathrm{\Lambda2(b)}}(p)\sim\frac{1}{2}\Lambda^{2}$,
the only quadratically divergent term and\begin{equation}
\Pi_{\mu\nu}^{\mathrm{\Lambda2}}(p)=\frac{i}{16\pi^{2}}\kappa^{2}\left(p^{2}\eta_{\mu\nu}-p_{\mu}p_{\nu}\right)\left(\frac{1}{2}\Lambda^{2}-\frac{1}{6}p^{2}\left(\ln\left(\frac{\Lambda^{2}}{p^{2}}\right)-\frac{5}{6}\right)\right).\label{eq:L2}\end{equation}
Notice that this result differs from \eqref{eq:L1} only in the value
and the sign of the coefficient of the first term, the change originates
from the different treatment of the graviton propagator.

The result of the improved momentum cutoff can be reproduced applying
dimensional regularization with care. The improved cutoff method works
in four physical dimensions and special rules have to be applied only
at the evaluation of the last tensor integrals. It is equivalent to
setting $d=4$ in the Einstein-Maxwell theory, e.g. both in the graviton
propagator and in the trace of the metric tensor. Dimensional regularization
is then applied at the last step evaluating the tensor and scalar
momentum integrals. We have found that $\Pi_{\mu\nu}^{\mathrm{DR3(b)}}=0$
and 

\begin{equation}
\Pi_{\mu\nu}^{DR3}(p)=\frac{i}{16\pi^{2}}\kappa^{2}\left(p^{2}\eta_{\mu\nu}-p_{\mu}p_{\nu}\right)\left(-\frac{1}{6}p^{2}\left(\frac{2}{\epsilon}+\ln\left(\frac{\mu^{2}}{p^{2}}\right)+\frac{5}{3}\right)\right).\label{eq:DR3}\end{equation}
 The corresponding cutoff result diverges quadratically and agrees
with the improved cutoff calculation (\ref{eq:impi1},\ref{eq:impi2})

\begin{eqnarray}
\Pi_{\mu\nu}^{\mathrm{\Lambda3}(a)}(p) & = & \frac{i}{16\pi^{2}}\kappa^{2}\left(p^{2}\eta_{\mu\nu}-p_{\mu}p_{\nu}\right)\left(-2\Lambda^{2}-\frac{1}{6}p^{2}\left(\ln\left(\frac{\Lambda^{2}}{p^{2}}\right)+\frac{2}{3}\right)\right),\label{eq:DR3i1}\\
\Pi_{\mu\nu}^{\mathrm{\Lambda3}(b)}(p) & = & \frac{i}{16\pi^{2}}\kappa^{2}\left(p^{2}\eta_{\mu\nu}-p_{\mu}p_{\nu}\right)\left(\phantom{2}\frac{3}{2}\Lambda^{2}\right).\label{eq:DR3i2}\end{eqnarray}

The quadratic divergences $\left(\Lambda^{2}\right)$ here are identified
with the $d=2$ poles in the extension \cite{veltman} of dimensional
regularizations \cite{dreg,leibb}. There may appear an additional
pole $1/(d-2)$ in the graviton propagator \eqref{eq:propg}. It is
coming from a non-physical point of the Einstein-Hilbert theory as
this theory is not a dynamical one in $d=2$, the Lagrangian reduces
to a trivial surface integral. In the first case, in \eqref{eq:L1}
we apply continuous $d$ both in the propagator \eqref{eq:propg}
and in the vertices during tracing. The second treatment sets $d=4$
in the propagator (as usually done in the literature) while using
continuous $d$ during tracing the indices. This hybrid treatment
looks not fully consistent as even in the loops one part of the theory
feels the modified $d$ dimensions the other part not, e.g. feels
fixed number of dimensions $d=4$ and gives \eqref{eq:DR2}. We prefer
the third, conceptionally simple case, when the gravity algebra is
performed in fixed $d=4$ and the rest of the calculation is done
using the standard dimensional regularization technique. Moreover,
the third result \eqref{eq:DR3i1} and \eqref{eq:DR3i2} agrees completely
with the improved cutoff calculation case.

In principle a theory is completely defined via specifying the Lagrangian
and the method of calculation e.g. fixing the regularization and the
treatment of the divergent terms, though the physical quantities must
be independent of the details of the regularization scheme. It is
remarkable that the transverse structure of the photon propagator
is not violated in any of the previous schemes and the logarithmic
term is universal in the three cases and agrees with earlier results
\cite{rodigast,deser}. The question is whether the $\Lambda^{2}$
terms contribute to the running of the gauge coupling, or have any
other effects on measurable physical quantities.

\section{Quadratic divergences and renormalization}

In the previous section we have calculated the 1-loop radiative correction
to the photon self energy from the effective theory of gravity in
the simplest Maxwell-Einstein theory. We have found under various
assumptions various quadratically divergent contributions (vanishing
particularly using a naive momentum cutoff). The 1-loop corrections
to the 2-point function generally modify the bare Lagrangian, the
divergences have to be removed by the properly chosen counterterms
via renormalization conditions. 

Consider the QED action with the convention \cite{don2}\begin{equation}
L_{0}=-\frac{1}{4e_{0}^{2}}F_{\mu\nu}F^{\mu\nu}+\bar{\Psi}iD_{\mu}\gamma^{\mu}\Psi,\qquad D_{\mu}=\partial_{\mu}+iA_{\mu}.\label{eq:Lqed}\end{equation}
The divergences calculated from the interaction \eqref{eq:EH} gives
the 1-loop effective action, here we focus only on the gravitational,
divergent contributions\begin{equation}
L=-\frac{1+a\kappa^{2}\Lambda^{2}}{4e_{0}^{2}}F_{\mu\nu}F^{\mu\nu}+a_{2}\ln\frac{\Lambda^{2}}{p^{2}}\left(D_{\mu}F^{\mu\nu}\right)^{2}+\left(\bar{\Psi}iD_{\mu}\gamma^{\mu}\Psi\right),\label{eq:LqedR}\end{equation}
where $p^{2}$ is the Euclidean momentum at which the 2-point function
was calculated. The question is whether should we interpret the coefficient
of the usual kinetic term as a varying, i.e. running electric charge
$\left(e^{2}(\Lambda)\simeq e_{0}^{2}\left(1-a\kappa^{2}\Lambda^{2}\right)\right)$?
The answer is no, because of the necessary wavefunction and charge
renormalization.

In quantum field theories the divergent terms have to be canceled
by the counterterms. New dimension-six term must be added to match
the $p^{2}\ln\left(\frac{\Lambda^{2}}{p^{2}}\right)$ term already
shown in \eqref{eq:LqedR}, \begin{equation}
L_{\mathrm{\mathrm{ct}}}=\frac{\delta Z_{1}}{4e_{0}^{2}}F_{\mu\nu}F^{\mu\nu}+\delta Z_{2}\left(D_{\mu}F^{\mu\nu}\right)^{2}.\label{eq:ct}\end{equation}
In principle there are three possible dimension-six counterterms $\left(D_{\mu}F^{\mu\nu}\right)^{2}$,
$\left(D_{\mu}F_{\nu\rho}\right)^{2}$ and $F_{\mu}^{\nu}F_{\nu}^{\rho}F_{\rho}^{\mu}$.
Only two of them are linearly independent up to total derivatives
and it turns out that the first, the $\left(D_{\mu}F^{\mu\nu}\right)^{2}$
term can cancel all divergences \cite{rodigast}. The coefficient
of the first term in \eqref{eq:LqedR} cannot be understood as defining
a running coupling but it is compensated by a counterterm through
a renormalization condition. It can be fixed either by the Coulomb
potential or Thomson scattering at low energy identifying the usual
electric charge as\begin{equation}
\frac{e_{0}^{2}}{4\pi(1+a\kappa^{2}\Lambda^{2})}=\frac{e^{2}}{4\pi}\simeq\frac{1}{137}.\label{eq: alpha}\end{equation}
Thus the quadratically divergent correction defines the relation between
the bare charge $e_{0}(\Lambda)$ in a theory with the physical cutoff
$\Lambda$ and the physical charge effective at low energies. After
fixing the parameters of the theory (e.g. by a measurement at low
energy) and using $e$ to calculate the predictions of the model the
cutoff dependence completely disappears from the physical charge \cite{don1,don2}.
The role of the quadratic correction is to define the relation \eqref{eq: alpha}
this way renormalizing the bare coupling constant $e_{0}(\Lambda)$
( and does not appear in the running of the physical charge).

Quadratic divergences are the main cause of the hierarchy problem
and discussed with other regularization methods. In \cite{guises}
the authors use Implicit Regularization, a general parametrization
of the basic divergent integrals, which separates the divergences
for a given problem in a process-independent way without referring
to a specific regularization (see also the Appendix). They argue that
their basic divergent integrals, thus the quadratic divergences can
be absorbed in the renormalization constants without explicitly determining
their value. Arbitrary parameters, such as the isolated quadratically
divergent contribution to the Higgs mass can be fixed by additional
(in the Higgs case: conformal) symmetry. Similar conclusion is reached
in \cite{afraid} using Wilsonian renormalization group (RG). They
argued that the additive (they call it subtractive) and multiplicative
renormalization procedure and the corresponding quadratic and logarithmic
divergences can be treated independently. They show that quadratic
divergences are the artifact of the regularization procedure and in
the Wilsonian RG they are naturally subtracted and simply define position
of the critical surface in the theory space. It is in complete agreement
with our claim in \eqref{eq: alpha} that the quadratic divergence
disappears from the physical quantities. The fate of the logarithmic
divergence could have been different.

The logarithmically divergent contribution on the other hand defines
the renormalization of the higher dimensional operator $\left(D_{\mu}F^{\mu\nu}\right)^{2}$
and again not the running of the gauge coupling. After renormalization
(at a point $p^{2}=\mu^{2}$) the logarithmic coefficient of the dim-6
term in \eqref{eq:LqedR} changes to $a_{2}\ln\frac{\Lambda^{2}}{p^{2}}-a_{2}\ln\frac{\Lambda^{2}}{\mu^{2}}=-a_{2}\ln\frac{p^{2}}{\mu^{2}}$
defining a would be running parameter. Furthermore note that this
term can be removed \cite{rodigast,ellis} by local field redefinition
of $A_{\mu}$ up to higher dimensional operators\begin{equation}
A_{\mu}\rightarrow A_{\mu}-c\nabla_{\nu}F_{\mu}^{\nu},\label{eq:redef}\end{equation}
 where $\nabla_{\mu}$ is the gravitational covariant derivative,
as the new term is proportional to the tree level equation of motion\begin{equation}
\nabla_{\mu}F^{\mu\nu}=0.\label{eq:29}\end{equation}
The logarithmic corrections were found in the first papers discussing
the gravitational contributions by Deser et al. \cite{deser} using
dimensional regularization and this way neglecting the quadratically
divergent contribution spotted by \cite{robinson}. Generally it can
be shown, that all photon propagator corrections can be removed by
appropriate field redefinition which are bilinear in $A_{\mu}$ even
if they contain arbitrary number of derivatives, on-shell scattering
processes are not influenced by the presence of such effective terms
\cite{ts}.

\section{Corrections to the gauge coupling in Yang-Mills theories}

We have discussed the simplest example including gravitational corrections
in Chapter 2, but already Deser, Tsao and Nieuenhuizen \cite{deser}
later Robinson and Wilczek \cite{robinson} and many other authors
performed their calculation in the Einstein-Yang-Mills system. Here
we follow the presentation of \cite{rodigast} to show that after
renormalization no meaningful running coupling can be defined even
identifying quadratic divergences using cutoff regularization.

Consider the Einstein-Yang-Mills Lagrangian

\begin{equation}
S=\int\mathrm{d}^{4}x\sqrt{-g}\left[\frac{2}{\kappa^{2}}R-\frac{1}{2}g^{\mu\nu}g^{\alpha\beta}\mathrm{Tr}\left[F_{\mu\nu}F_{\alpha\beta}\right]\right],\label{eq:EYM}\end{equation}
where the field strength has a Yang-Mills index $F_{\mu\nu}=\partial_{\mu}A_{\nu}-\partial_{\nu}A_{\mu}-ig\left[A_{\mu},\: A_{\nu}\right]$
and $g$ is the Yang-Mills coupling.

In the Yang-Mills theory the bare gluon three-point functions get
modified by gravity, too. Beside the corrections to the gluon two
point functions, which are order $\kappa^{2}$ and are the same as
presented in Chapter 2 see \eqref{eq:cupi1} and \eqref{eq:cupi2}
and Figure 1, there are contributions to the gluon three-point functions
at the $g\kappa^{2}$ order.

\begin{figure}
\begin{centering}
(c)\includegraphics[height=2.5cm]{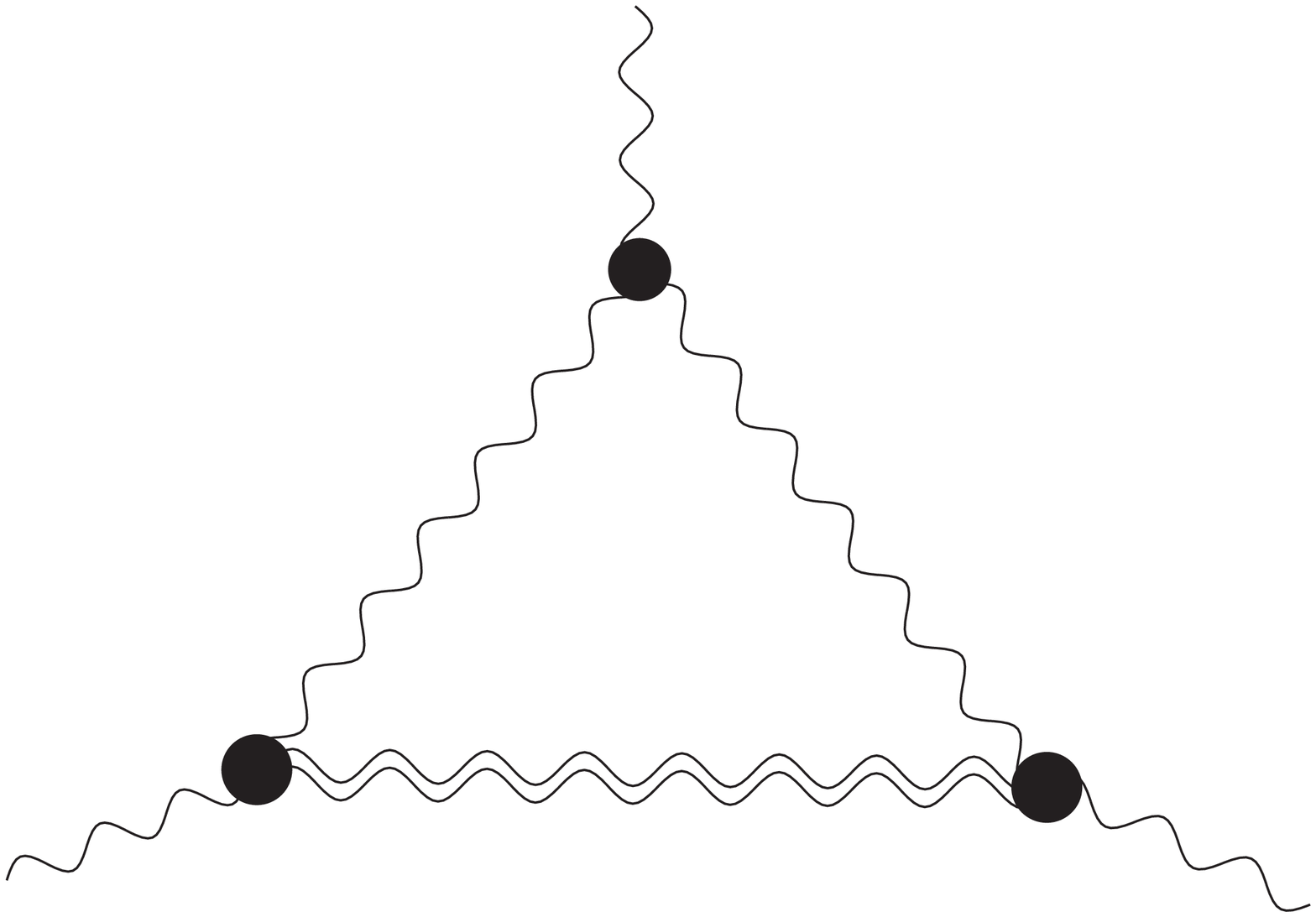} (d)\includegraphics[height=2.5cm]{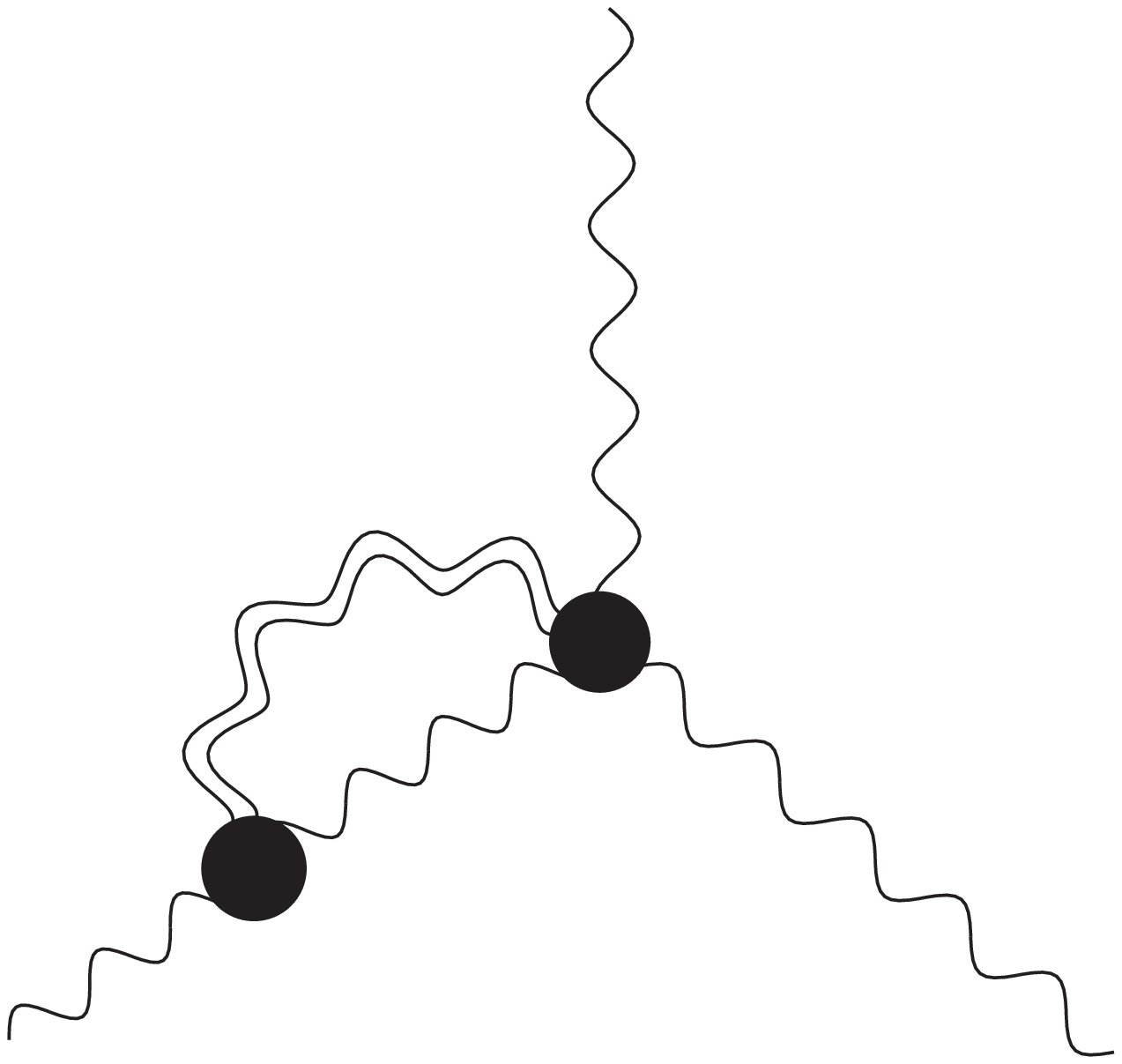}
(e)\includegraphics[height=2.5cm]{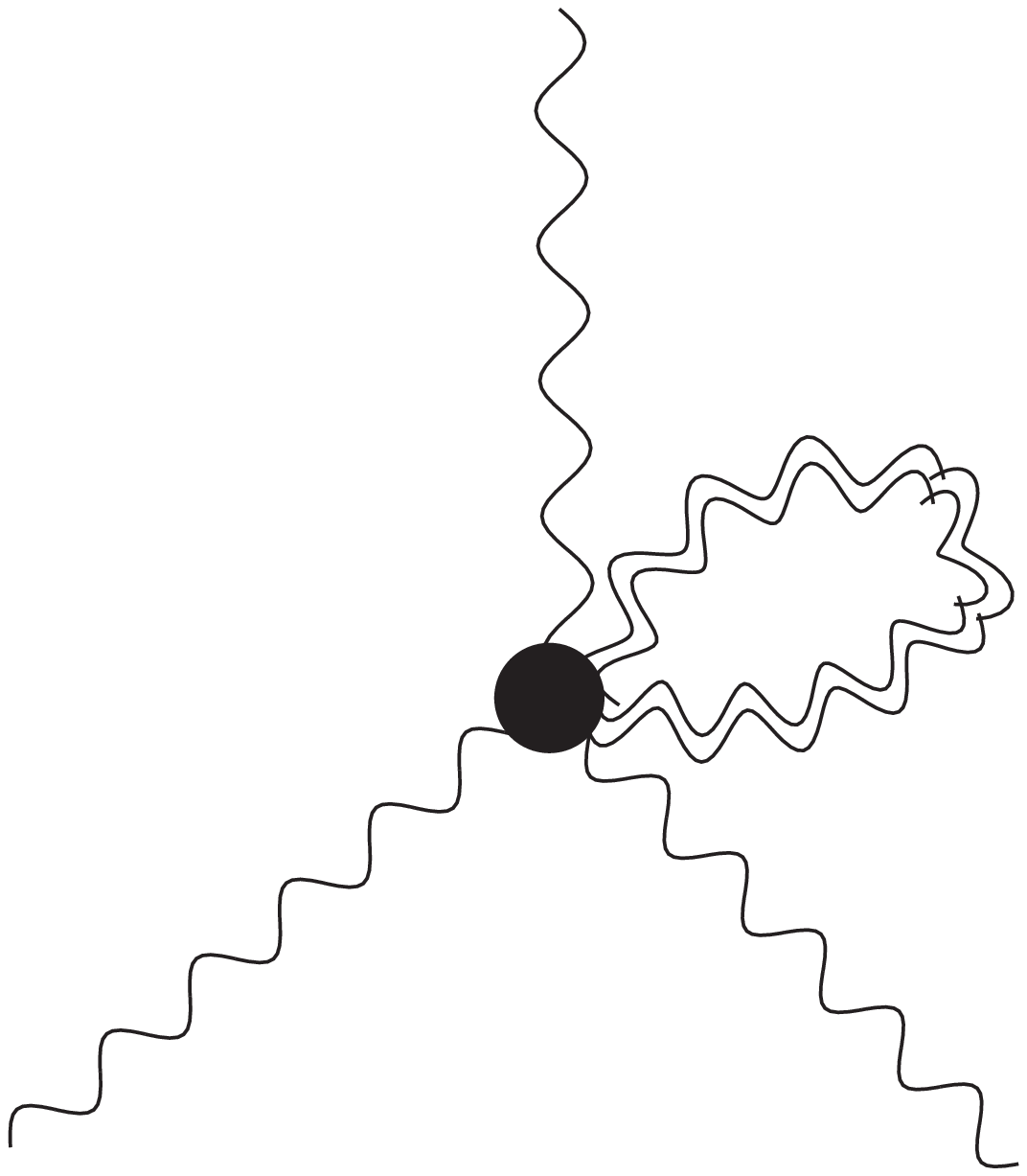}
\par\end{centering}

\centering{}\caption{Feynman graphs with graviton (double) lines contributing to the gluon
three point function.}

\end{figure}

There are new vertices with three selfinteracting gluon with extra
one- and two-graviton legs. The three gluon-one graviton vertex is

{\small \begin{eqnarray}
V_{gggG}(k_{1\mu}^{(a)},k_{2\nu}^{(b)},k_{3\rho}^{(c)},\alpha\beta) & \!=\! & -\! ig\kappa f^{abc}\left[P^{\alpha\beta,\mu\nu}\left(k_{1}-k_{2}\right)^{\rho}+\eta^{\alpha\beta}\left(\eta^{\rho\alpha}\left(k_{1}-k_{2}\right)^{\beta}+\eta^{\rho\alpha}\left(k_{1}-k_{2}\right)^{\beta}\right)+\right.\nonumber \\
 &  & \!\left.+\mathrm{cycl.perm.}\left\{ (\mu,k_{1}),(\nu,k_{2}),(\rho,k_{3})\right\} \right],\label{eq:V3g}\end{eqnarray}
}where $f^{abc}$ is the Yang-Mills structure constant. The three
gluon-two graviton vertex is again rather complicated and lengthy

\begin{eqnarray}
V_{gggGG}(k_{1\mu}^{(a)},k_{2\nu}^{(b)},k_{3\rho}^{(c)},\alpha\beta,\gamma\delta) & \!=\! & -ig\kappa^{2}f^{abc}\left[\left(k_{1}-k_{2}\right)^{\rho}\left(I^{\mu\nu,\alpha\gamma}\eta^{\delta\beta}+I^{\mu\nu,\alpha\delta}\eta^{\gamma\beta}+\left\{ ^{(\mu\nu)\longleftrightarrow(\alpha\beta)}\right\} \right.\right.\nonumber \\
 &  & -\left.\frac{1}{2}\left(\eta^{\alpha\beta}I^{\mu\nu,\gamma\delta}+\eta^{\gamma\delta}I^{\mu u,\alpha\beta}\right)-\eta^{\mu\nu}P^{\alpha\beta,\gamma\delta}\right)\nonumber \\
 &  & +\left(2\eta^{\mu\nu}P^{\gamma\delta,\alpha\beta}+I^{\mu\nu,\gamma\delta}\right)\left(k_{1}-k_{2}\right)^{\rho}+\left\{ ^{(\alpha)\longleftrightarrow(\beta)}\right\} \nonumber \\
 &  & \!\left.\left\{ ^{(\gamma\delta)\longleftrightarrow(\alpha\beta)}\right\} +\mathrm{cycl.perm.}\left\{ (\mu,k_{1}),(\nu,k_{2}),(\rho,k_{3})\right\} \right].\label{eq:V3gg}\end{eqnarray}
With these vertices there are three graphs contributing to the gluon
three-point function at one-loop, Fig.2. The external gluons are labeled
as in the vertices $\left\{ (\mu,k_{1}),(\nu,k_{2}),(\rho,k_{3})\right\} $
and the the 3-point function contributions must be symmetrized in
these index-pairs. The graph $(c)$ is only logarithmically divergent\begin{equation}
G_{3}^{(c)}\sim\frac{1}{16\pi^{2}}g\kappa^{2}f^{abc}\log\Lambda^{2}F_{3}^{\mu\nu\rho}\left(k_{1}^{\mu},k_{2}^{\nu},k_{3}^{\rho}\right),\label{eq:g3c}\end{equation}
where the lengthy $F_{3}^{\mu\nu\rho}$ function scales with the third
power of momenta. The graphs (d) and (e) are similar to (a) and (b)
only with the exception of an additional gluon leg starting from the
main vertex. Graph (d) has similar logarithmic correction as \eqref{eq:g3c}
and a quadratic divergence, while in (e) the divergence is purely
quadratic.\begin{eqnarray}
G_{3}^{(d)} & \simeq & \phantom{{\}}}\frac{1}{16\pi^{2}}g\kappa^{2}f^{abc}\frac{3}{2}\left(\eta^{\mu\nu}\left(k_{1}-k_{2}\right)^{\rho}+\mathrm{symmetrized}\right)\Lambda^{2}+\mathrm{log\: terms},\label{eq:g3d}\\
G_{3}^{(e)} & \simeq & -\frac{1}{16\pi^{2}}g\kappa^{2}f^{abc}\frac{3}{2}\left(\eta^{\mu\nu}\left(k_{1}-k_{2}\right)^{\rho}+\mathrm{symmetrized}\right)\Lambda^{2}.\end{eqnarray}
The sum of the quadratic contributions from graphs (d) and (e) exactly
cancel just as for the two point functions in Fig. 1. The remaining
logarithmic divergence surprisingly can be canceled by only the second
term in \eqref{eq:ct}.\textcompwordmark{}\begin{equation}
L_{c.t.}\supset\frac{1}{16\pi^{2}}\frac{1}{6}\kappa^{2}\log\frac{\Lambda^{2}}{\mu^{2}}\left(D_{\mu}F^{\mu\nu}\right),\label{eq:ct2}\end{equation}
where $\mu$ is the renormalization scale in agreement with the result
of \cite{deser} and later works. We emphasize that the counterterm
in \eqref{eq:ct2} corrects a higher dimensional operator, and the
contribution can be removed by a non-linear field redefinitions of
the gauge field \eqref{eq:redef} as discussed in Chapter 3 and does
not lead to a change in the running of physical parameters.

\section{Conclusion}

We have calculated and presented in this chapter the gravitational
corrections to gauge theories in the framework of effective field
theories. The study was motivated by the various, sometime controversial
results in the literature. Our method and the presented results were
capable of identifying quadratically divergent contributions to the
photon and generalized gluon two and three point functions, thanks
to the gauge invariant construction. In the first, QED part, to test
our calculation we defined the cutoff dependence employing \eqref{eq: quad},
\eqref{eq:log} and dimensional regularization with various assumptions
about treating the number of dimensions $d$. We observed that the
1-loop gravity corrections to the two point function in all but one
cases contain $\Lambda^{2}$ divergence with the exception of the
naive momentum cutoff which violates gauge symmetries usually. Here
all the corrections are transverse. The logarithmic term universally
agrees with the literature starting from Deser et al. \cite{deser}.
Then we presented the corrections in a more general Yang-Mills theory.
We found that the logarithmically divergent terms contribute to the
dimension-6 terms and can be removed by local field redefinitions
this way do not affecting the running of the gauge coupling. $\Lambda^{2}$
corrections to the QED or Yang-Mills effective actions are absent
using a naive cutoff regularizations and are present with more sophisticated
methods, but those are proved to be non-physical.

The quadratically divergent corrections to the photon or gluon self-energy
do not lead to the modification of the running of the gauge coupling.
Robinson and Wilczek claimed that the $-a\kappa^{2}\Lambda^{2}$ correction
could turn the beta function negative and make the Einstein-Maxwell
and Einstein-Yang-Mills theory asymptotically free. This statement
and the calculation was criticized in the literature. We showed in
this chapter using explicit cutoff calculation that $\Lambda^{2}$
corrections may appear in the 2-point function, but those will define
the renormalization connection between the cutoff dependent bare coupling
and the physical coupling \eqref{eq: alpha} and do not lead to a
running coupling. This conclusion is in complete agreement with other
results concerning quadratic divergences \cite{don1,guises,afraid}.
Indeed the $\Lambda^{2}$ correction can be absorbed into the physical
charge and does not appear in physical processes. Donoghue et al.
argue in \cite{don1} that an universal, i.e. process independent
running coupling constant cannot be defined in the effective theory
of gravity independently of the applied regularization. They demonstrate
that because of the crossing symmetry in theories (except the $\lambda\Phi^{4}$)
even the sign of the would be quadratic running is ambiguous and a
running coupling would be process dependent, thus not useful. Generally
the logarithmically divergent corrections could define the renormalization
of higher dimensional operators. It turns out that even these logarithmic
correction can be removed by appropriate field redefinitions and do
not contribute to on-shell scattering processes. We note that the
authors in \cite{brazil} showed using their 4-dimensional implicit
regularization method that the quadratic terms are coming from ambiguous
surface terms, discussed in more details in \cite{uj,sampaio}, and
as such are non-physical. Interestingly those surface terms vanish
if we evaluate them with our improved cutoff \cite{nova15}.

Finally we point out that we have found gravity corrections to the
two and three-point functions in gauge theories. Using a momentum
cutoff the quadratically divergent contributions define the renormalization
of the bare charge and thus using the physical charge the $\Lambda^{2}$
corrections do not appear in physical processes. On the other hand
logarithmic corrections are universal but merely define the renormalization
of a dimension-6 term in the Lagrangian, which term can be eliminated
by local field redefinition. We conclude that gravity corrections
do not lead to the modification of the usual running of gauge coupling
and cannot point towards asymptotic freedom in the case of gauge theories.

\part*{Appendix: Improved momentum cutoff}

In this appendix we introduce the novel regularization of gauge theories,
proposed in \cite{uj} and discussed with broader outlook on the literature
in \cite{nova15}. It is based on 4 dimensional momentum cutoff to
evaluate 1-loop divergent integrals. The idea was to construct a cutoff
regularization which does not brake gauge symmetries and the necessary
shift of the loop-momentum is allowed as no surface terms are generated.
The loop calculation starts with Wick rotation, Feynman-parametrization
and loop-momentum shift. Only the treatment of free Lorentz indices
under divergent integrals should be changed compared to the naive
cutoff calculation.

We start with the observation that the contraction with $\eta_{\mu\nu}$
(tracing) does not necessarily commute with loop-integration in divergent
cases. Therefore the substitution of \begin{equation}
k_{\mu}k_{\nu}\rightarrow\frac{1}{4}\eta_{\mu\nu}k^{2}\label{eq:negyed}\end{equation}
is not valid under divergent integrals, where $k$ is the loop-momentum%
\footnote{The metric tensor is denoted by $\eta_{\mu\nu}$ both in Minkowski
and Euclidean space. %
}. The usual factor $1/4$ is the result of tracing both sides under
the loop integral, e.g. changing the order of tracing and the integration.
In the new approach the integrals with free Lorentz indices are defined
using physical consistency conditions, such as gauge invariance or
freedom of momentum routing. Based on the diagrammatic proof of gauge
invariance it can be shown that the two conditions are related and
both are in connection with the requirement of vanishing surface terms.
It was proposed in \cite{uj} that instead of \eqref{eq:negyed} the
general identification of the cutoff regulated integrals in gauge
theories \begin{equation}
\int_{\Lambda\: reg}d^{4}l_{E}\frac{l_{E\mu}l_{E\nu}}{\left(l_{E}^{2}+m^{2}\right)^{n+1}}=\frac{1}{2n}\eta_{\mu\nu}\int_{\Lambda\: reg}d^{4}l_{E}\frac{1}{\left(l_{E}^{2}+m^{2}\right)^{n}},\ \ \ \ \ n=1,2,...\label{eq:idn}\end{equation}
will satisfy the Ward-Takahashi identities and gauge invariance at
1-loop ($l{}_{E}$ is the shifted Euclidean loop-momentum). In case
of divergent integrals it differs from \eqref{eq:negyed}, for non-divergent
cases both substitutions give the same results at ${\cal O}(1/\Lambda^{2})$
(the difference is a vanishing surface term). It is shown in \cite{uj}
that this definition is robust in gauge theories, differently organized
calculations of the 1-loop functions agree with each other using \eqref{eq:idn}
and disagree using \eqref{eq:negyed}. For four free indices the gauge
invariance dictates ($n=2,3,...$)\begin{equation}
\int_{\Lambda\: reg}d^{4}l_{E}\frac{l_{E\alpha}l_{E\beta}l_{E\mu}l_{E\rho}}{\left(l_{E}^{2}+m^{2}\right)^{n+1}}=\frac{1}{4n(n-1)}\int_{\Lambda\: reg}d^{4}l_{E}\frac{\eta_{\alpha\beta}\eta_{\mu\rho}+\eta_{\alpha\mu}\eta_{\beta\rho}+\eta_{\alpha\rho}\eta_{\beta\mu}}{\left(l_{E}^{2}+m^{2}\right)^{n-1}}.\label{eq:idn4}\end{equation}
For 6 and more free indices appropriate rules can be derived (or \eqref{eq:idn}
can be used recursively for each allowed pair). Finally the scalar
integrals are evaluated with a simple Euclidean momentum cutoff. The
method was successfully applied to an effective model to estimate
oblique corrections \cite{fcmlambda}.

There are similar attempts to define a regularization that respects
the original gauge and Lorentz symmetries of the Lagrangian but work
in four spacetime dimensions usually with a cutoff \cite{gu,wu1}.
Some methods can separate the divergences of the theories and does
not rely on a physical cutoff \cite{nemes,rosten,polon} or even could
be independent of it \cite{pittau}. For further literature see references
in \cite{uj}.

Under this modified cutoff regularization the terms with numerators
proportional to the loop momentum are all defined by the possible
tensor structures. Odd number of $l_{E}$'s give zero as usual, but
the integral of even number of $l_{E}$'s is defined by \eqref{eq:idn},
\eqref{eq:idn4} and similarly for more indices, this guarantees that
the symmetries are not violated. The calculation is performed in 4
dimensions, the finite terms are equivalent with the results of dimensional
regularization. The method identifies quadratic divergences while
gauge and Lorentz symmetries are respected. We stress that the method
treats differently momenta with free ($k_{\mu}k_{\nu}$) and contracted
Lorentz indices ($k^{2}$), the order of tracing and performing the
regulated integral cannot be changed similarly to dimensional regularization.
The famous triangle anomaly can be unambiguously defined and presented
in \cite{tranom} see also \cite{white}, \cite{gam5b}.

However even using dimensional regularization one is able to define
cutoff results in agreement with the present method. In dimensional
regularization singularities are identified as $1/\epsilon$ poles,
power counting shows that these are the logarithmic divergences of
the theory. Naively quadratic divergences are set to zero in the process,
but already Veltman noticed \cite{veltman} that these divergences
can be identified by calculating the poles in $d=2$ $(\epsilon=1)$.
Careful calculation of the Veltman-Passarino 1-loop functions in dimensional
regularization and with 4-momentum cutoff leads to the following identifications
\cite{uj,hagiwara,harada} \begin{eqnarray}
4\pi\mu^{2}\left(\frac{1}{\epsilon-1}+1\right) & = & \Lambda^{2},\label{eq: quad}\\
\frac{1}{\epsilon}-\gamma_{E}+\ln\left(4\pi\mu^{2}\right)+1 & = & \ln\Lambda^{2}.\label{eq:log}\end{eqnarray}
The finite terms are unambiguously defined \begin{equation}
f_{{\rm finite}}=\lim_{\epsilon\rightarrow0}\left[f(\epsilon)-R(0)\left(\frac{1}{\epsilon}-\gamma_{E}+\ln4\pi+1\right)-R(1)\left(\frac{1}{\epsilon-1}+1\right)\right],\label{eq:finite}\end{equation}
where $R(0)$, \, $R(1)$ are the residues of the poles at $\epsilon=0,\,1$
respectively. Using \eqref{eq: quad}, \eqref{eq:log} and \eqref{eq:finite}
at 1-loop the results of the improved cutoff can be reproduced using
dimensional regularization without any ambiguous subtraction.

The loop integrals are calculated as follows. First the loop momentum
($k$) integral is Wick rotated (to $k_{E}$), with Feynman parameter(s)
the denominators are combined, then the order of Feynman parameter
and the momentum integrals are changed. After that the loop momentum
($k_{E}\rightarrow l_{E}$) is shifted to have a spherically symmetric
denominator.

Finally we present two divergent integrals calculated by the new regularization.
$\Delta$ can be any loop momentum independent expression depending
on the Feynman $x$ parameter, external momenta, masses, e.g. $\Delta(x,q_{i},m).$
The integration is understood for Euclidean momenta with absolute
value below the $\Lambda$ cutoff $\left(|l_{E}|\leq\Lambda\right)$.

The integral \eqref{eq:A1} is just given for comparison, it is calculated
with a simple momentum cutoff. In \eqref{eq:A2} with the standard
\eqref{eq:negyed} substitution one would get a constant $-\frac{3}{2}$
instead of $-1$ \cite{uj}.

\begin{eqnarray}
\int_{\Lambda\, reg}\frac{d^{4}l_{E}}{i(2\pi)^{4}}\frac{1}{\left(l_{E}^{2}+\Delta^{2}\right)^{2}} & \!\!=\!\! & \frac{1}{(4\pi)^{2}}\left(\ln\left(\frac{\Lambda^{2}+\Delta^{2}}{\Delta^{2}}\right)+\frac{\Delta^{2}}{\Lambda^{2}+\Delta^{2}}-1\right).\label{eq:A1}\\
\int_{\Lambda\, reg}\frac{d^{4}k}{i(2\pi)^{4}}\frac{l_{E\mu}l_{E\nu}}{\left(l_{E}^{2}+\Delta^{2}\right)^{3}} & \!\!=\!\! & \frac{1}{(4\pi)^{2}}\frac{g_{\mu\nu}}{4}\left(\ln\left(\frac{\Lambda^{2}+\Delta^{2}}{\Delta^{2}}\right)+\frac{\Delta^{2}}{\Lambda^{2}+\Delta^{2}}-1\right).\label{eq:A2}\end{eqnarray}


\begin{thebibliography}{52}
\bibitem{ligo}B.~P.~Abbott {\it et al.} [LIGO Scientific and Virgo Collaborations],   %``Observation of Gravitational Waves from a Binary Black Hole Merger,''   
Phys.\ Rev.\ Lett.\  {\bf 116} (2016) no.6,  061102.

\bibitem{veltmanh}G.~'t Hooft and M.~J.~G.~Veltman,   %``One loop divergencies in the theory of gravitation,''   
Annales Poincare Phys.\ Theor.\ A {\bf 20} (1974) 69.

\bibitem{greff} J.~F.~Donoghue,   %``General relativity as an effective field theory: The leading quantum corrections,''   
Phys.\ Rev.\ D {\bf 50} (1994) 3874. 
%   [gr-qc/9405057]. 

\bibitem{greff2}J.~F.~Donoghue,   %``The effective field theory treatment of quantum gravity,''  
AIP Conf.\ Proc.\  {\bf 1483} (2012) 73.  % [arXiv:1209.3511 [gr-qc]].

\bibitem{burg}C.~P.~Burgess,   %``Quantum gravity in everyday life: General relativity as an effective field theory,''   
Living Rev.\ Rel.\  {\bf 7} (2004) 5. %   [gr-qc/0311082].

\bibitem{atlas}  G.~Aad {\it et al.}  [ATLAS Collaboration],   %``Observation of a new particle in the search for the Standard Model Higgs boson with the ATLAS detector at the LHC,''   
Phys.\ Lett.\ B {\bf 716} (2012) 1.  % [arXiv:1207.7214 [hep-ex]].   

\bibitem{cms}S.~Chatrchyan {\it et al.}  [CMS Collaboration],   
%``Observation of a new boson at a mass of 125 GeV with the CMS experiment at the LHC,''   
Phys.\ Lett.\ B {\bf 716} (2012) 30.  % [arXiv:1207.7235 [hep-ex]].

\bibitem{stabil} G.~Degrassi, S.~Di Vita, J.~Elias-Miro, J.~R.~Espinosa, G.~F.~Giudice, G.~Isidori and A.~Strumia,   %``Higgs mass and vacuum stability in the Standard Model at NNLO,''   
JHEP {\bf 1208} (2012) 098. %   [arXiv:1205.6497 [hep-ph]].   
%%CITATION = ARXIV:1205.6497;%%

\bibitem{misa}F.~Bezrukov, M.~Y.~Kalmykov, B.~A.~Kniehl and M.~Shaposhnikov,   %``Higgs Boson Mass and New Physics,''
JHEP {\bf 1210} (2012) 140.  % [arXiv:1205.2893 [hep-ph]].

\bibitem{stab3}D.~Buttazzo, G.~Degrassi, P.~P.~Giardino, G.~F.~Giudice, F.~Sala, A.~Salvio and A.~Strumia,   %``Investigating the near-criticality of the Higgs boson,''
arXiv:1307.3536 [hep-ph].

\bibitem{deser} S.~Deser, H.~-S.~Tsao and P.~van Nieuwenhuizen,   %``One Loop divergencies of the Einstein Yang-Mills System,''   
Phys.\ Rev.\ D {\bf 10} (1974) 3337.

\bibitem{robinson}S.~P.~Robinson and F.~Wilczek, 
%``Gravitational correction to running of gauge couplings,'' 
Phys.\ Rev.\ Lett.\ {\bf 96} (2006) 231601. % [hep-th/0509050]. %%CITATION = HEP-TH/0509050;%%

\bibitem{pietr}A.~R.~Pietrykowski,   %``Gauge dependence of gravitational correction to running of gauge couplings,''  
Phys.\ Rev.\ Lett.\  {\bf 98} (2007) 061801.%   [hep-th/0606208].

\bibitem{toms1} D.~J.~Toms,   %``Quantum gravity and charge renormalization,'' 
 Phys.\ Rev.\ D {\bf 76} (2007) 045015. %   [arXiv:0708.2990 [hep-th]].

\bibitem{rodigast}D.~Ebert, J.~Plefka and A.~Rodigast,
%``Absence of gravitational contributions to the running Yang-Mills coupling,''   
Phys.\ Lett.\ B {\bf 660} (2008) 579. %   [arXiv:0710.1002 [hep-th]].

\bibitem{toms2}D.~J.~Toms,   %``Quantum gravitational contributions to quantum electrodynamics,''
Nature {\bf 468} (2010) 56.  % [arXiv:1010.0793 [hep-th]].

\bibitem{wu}  Y.~Tang and Y.~-L.~Wu,   %``Quantum Gravitational Contributions to Gauge Field Theories,''   
Commun.\ Theor.\ Phys.\  {\bf 57} (2012) 629. 
%   [arXiv:1012.0626 [hep-ph]].

\bibitem{pietr2} A.~R.~Pietrykowski,  
Phys.\ Rev.\ D {\bf 87} (2013) 024026. %   [arXiv:1210.0507 [hep-th]].

\bibitem{he} H.~-J.~He, X.~-F.~Wang and Z.~-Z.~Xianyu,   %``Gauge-Invariant Quantum Gravity Corrections to Gauge Couplings via Vilkovisky-DeWitt Method and Gravity Assisted Gauge Unification,''   
Phys.\ Rev.\ D {\bf 83} (2011) 125014. %   [arXiv:1008.1839 [hep-th]].

\bibitem{nielsen}N.~K.~Nielsen,   %``The Einstein-Maxwell system, Ward identities, and the Vilkovisky construction,''   
Annals Phys.\  {\bf 327} (2012) 861. %   [arXiv:1109.2699 [hep-th]].

\bibitem{weinberg}S. Weinberg, in General Relativity: An Einstein Centenary Survey, ed. by S. Hawking and W. Israel, Cambridge UNiversity press, 1979, p.790

\bibitem{reuter}M.~Niedermaier and M.~Reuter,   %``The Asymptotic Safety Scenario in Quantum Gravity,''
Living Rev.\ Rel.\  {\bf 9} (2006) 5.

\bibitem{reuter2}J.~-E.~Daum, U.~Harst and M.~Reuter,   %``Running Gauge Coupling in Asymptotically Safe Quantum Gravity,''   
JHEP {\bf 1001} (2010) 084. %   [arXiv:0910.4938 [hep-th]]

\bibitem{litim}S.~Folkerts, D.~F.~Litim and J.~M.~Pawlowski,   
%``Asymptotic freedom of Yang-Mills theory with gravity,''   
Phys.\ Lett.\ B {\bf 709} (2012) 234. %   [arXiv:1101.5552 [hep-th]].

\bibitem{narain}G.~Narain and R.~Anishetty,   %``Charge Renormalization due to Graviton Loops,''   
JHEP {\bf 1307} (2013) 106. %  [arXiv:1211.5040 [hep-th]]

\bibitem{ellis} J.~Ellis and N.~E.~Mavromatos,   %``On the Interpretation of Gravitational Corrections to Gauge Couplings,'' 
 Phys.\ Lett.\ B {\bf 711} (2012) 139. %   [arXiv:1012.4353 [hep-th]].

\bibitem{don1}  M.~M.~Anber, J.~F.~Donoghue and M.~El-Houssieny,   %``Running couplings and operator mixing in the gravitational corrections to coupling constants,''   
Phys.\ Rev.\ D {\bf 83} (2011) 124003. %   [arXiv:1011.3229 [hep-th]].

\bibitem{brazil}J.~C.~C.~Felipe, L.~C.~T.~Brito, M.~Sampaio and M.~C.~Nemes,   %``Quantum gravitational contributions to the beta function of quantum electrodynamics,''
Phys.\ Lett.\ B {\bf 700} (2011).

\bibitem{grcutoff} G.~Cynolter and E.~Lendvai,   %``Corrections to gauge theories in effective quantum gravity with a cutoff,''   
Mod.\ Phys.\ Lett.\ A {\bf 29} (2014) 1450024.
%doi:10.1142/S0217732314500242   [arXiv:1307.4651 [hep-ph]].

\bibitem{uj}  G.~Cynolter and E.~Lendvai,   %``Symmetry preserving regularization with a cutoff,''
Central Eur.J.Phys. {\bf 9} (2011) 1237. 
%-1247  arXiv:1002.4490 [hep-ph]

\bibitem{form}J.~A.~M.~Vermaseren,   ``New features of FORM,''   math-ph/0010025.

\bibitem{don2}M.~M.~Anber and J.~F.~Donoghue, %``On the running of the gravitational constant,''
Phys.\ Rev.\ D {\bf 85} (2012) 104016. %   [arXiv:1111.2875 [hep-th]].

\bibitem{guises}A.~L.~Cherchiglia, A.~R.~Vieira, B.~Hiller, A.~P.~Baêta Scarpelli and M.~Sampaio,   %``Guises and Disguises of Quadratic Divergences,''   
Annals Phys.\  {\bf 351} (2014) 751   doi:10.1016/j.aop.2014.10.002

\bibitem{afraid}H.~Aoki and S.~Iso, %``Revisiting the Naturalness Problem -- Who is afraid of quadratic divergences? --,''
Phys.\ Rev.\ D {\bf 86} (2012) 013001   doi:10.1103/PhysRevD.86.013001

\bibitem{ts}A.~A.~Tseytlin,
%``Ambiguity in the Effective Action in String Theories,''   
Phys.\ Lett.\ B {\bf 176} (1986) 92.

\bibitem{nova15}G.~Cynolter and E.~Lendvai,   %``Cutoff Regularization Method in Gauge Theories,''
in Gauge Theories and Differential Geometry, (editor Lance Bailey), pp. 199-218  ISBN: 978-1-63483-546-6, arXiv:1509.07407 [hep-ph].

\bibitem{fcmlambda} G.~Cynolter, E.~Lendvai and G.~P\'ocsik,
 %``S and T Parameters in the Fermion Condensate Model,''
Mod.\ Phys.\ Lett.\  A {\bf 24} (2009) 2331.

\bibitem{gu}Y.~Gu,   
%``Momentum-cutoff regularization and gauge invariance in QED,''
J.\ Phys.\ A  {\bf 39} (2006) 13575.

\bibitem{wu1}Y.~L.~Wu,   
%``Symmetry principle preserving and infinity free regularization and   %renormalization of quantum field theories and the mass gap,''
Int.\ J.\ Mod.\ Phys.\  A {\bf 18} (2003) 5363.

\bibitem{nemes} O.~A.~Battistel and M.~C.~Nemes,   
%``Consistency in regularizations of the gauged NJL model at one loop   %level,''   
Phys.\ Rev.\  D {\bf 59} (1999) 055010.

\bibitem{rosten}S.~Arnone, T.~R.~Morris and O.~J.~Rosten,   
%``A Generalised Manifestly Gauge Invariant 
%Exact Renormalisation Group for   %SU(N) Yang-Mills,''   
Eur.\ Phys.\ J.\  C {\bf 50} (2007) 467.

\bibitem{polon} F.~del Aguila and M.~Perez-Victoria,   %``Differential renormalization of gauge theories,''   
Acta Phys.\ Polon.\  B {\bf 29} (1998) 2857.  % [arXiv:hep-ph/9808315].

\bibitem{sampaio} A.~R.~Vieira, A.~L.~Cherchiglia and M.~Sampaio,   %``Momentum Routing Invariance in Extended QED: Assuring Gauge Invariance Beyond Tree Level,''   
Phys.\ Rev.\ D {\bf 93} (2016) no.2,  025029   doi:10.1103/PhysRevD.93.025029

\bibitem{pittau}R.~Pittau,   %``A four-dimensional approach to quantum field theories,''
JHEP {\bf 1211} (2012) 151. 
%arXiv:1208.5457 [hep-ph].

\bibitem{tranom}G.~Cynolter and E.~Lendvai,   %``Note on triangle anomaly with improved momentum cutoff,''   
Mod.\ Phys.\ Lett.\ A {\bf 26} (2011) 1537.
% [arXiv:1012.4648 [hep-ph]].   %%CITATION = ARXIV:1012.4648;%%

\bibitem{white}B.~K.~El-Menoufi and G.~A.~White,   %``The axial anomaly, dimensional regularization and Lorentz-violating QED,''  
arXiv:1505.01754 [hep-th].

\bibitem{gam5b}A.~C.~D.~Viglioni, A.~L.~Cherchiglia, A.~R.~Vieira, B.~Hiller and M.~Sampaio,   %``$\gamma_{5}$ algebra ambiguities in Feynman amplitudes: momentum routing invariance and anomalies in $D=4$ and $D=2$,''   
arXiv:1606.01772 [hep-th].

\bibitem{veltman}M.J.G Veltman, Acta Phys. Polon. \textbf{B12}: (1981)
437.

\bibitem{hagiwara} K.~Hagiwara, S.~Ishihara, R.~Szalapski and
D.~Zeppenfeld, %``Low-energy effects of new interactions in the electroweak boson sector,''
 Phys.\ Rev.\ D \textbf{48} (1993) 2182.

\bibitem{harada}M.~Harada and K.~Yamawaki,   
%``Wilsonian matching of effective field theory with underlying QCD,''  
Phys.\ Rev.\  D {\bf 64}, 014023 (2001).   %[arXiv:hep-ph/0009163].

\bibitem{dreg}G.~'t Hooft and M.~J.~G.~Veltman,
%``Regularization And Renormalization Of Gauge Fields,''
Nucl.\ Phys.\  B {\bf 44} (1972) 189.

\bibitem{leibb}G.~Leibbrandt,   %``Introduction to the Technique of Dimensional Regularization,''
 Rev.\ Mod.\ Phys.\  {\bf 47} (1975) 849.
\end{thebibliography}
\end{document}